\documentclass[prc,aps,final,twocolumn,nofootinbib]{revtex4-1}
\usepackage{graphicx}
\usepackage{mathrsfs}
\usepackage{bm}
\usepackage{colordvi}

\newcommand{\bel}[1]{\begin{equation}\label{#1}}

\usepackage{epsfig,psfrag}

\def\ramaSM{\vadjust{\vbox to 0pt{\vss \hbox to \hsize
{\hskip\hsize \quad $\Leftarrow$\quad {\it SM}\hss}\vskip3.5pt}}}
\def\ramaSL{\vadjust{\vbox to 0pt{\vss \hbox to \hsize
{\hskip\hsize \quad $\Leftarrow$\quad {\it SL}\hss}\vskip3.5pt}}}
\def\ramaSR{\vadjust{\vbox to 0pt{\vss \hbox to \hsize
{\hskip\hsize \quad $\Leftarrow$\quad {\it SR}\hss}\vskip3.5pt}}}
\def\rama{\vadjust{\vbox to 0pt{\vss \hbox to \hsize
{\hskip\hsize \quad $\Leftarrow$\quad
{$\Longleftarrow$}\hss}\vskip3.5pt}}}

\def\be{\begin{equation}}
\def\ee{\end{equation}}
\def\bea{\begin{eqnarray}}
\def\eea{\end{eqnarray}}

\def\hahat{\hat{H}}

\def\hahat0{\hat{H}_0}

\def\exp{\hbox{exp}}

\def\siml{\hbox{\kern.1em \lower.6ex \hbox{$\sim$} \kern-1.12em
          \raise.6ex \hbox{$<$} \kern.1em }}
\def\simg{\hbox{\kern.1em \lower.6ex \hbox{$\sim$} \kern-1.12em
          \raise.6ex \hbox{$>$} \kern.1em }}

\def\siml{\hbox{\kern.1em \lower.6ex \hbox{$\sim$} \kern-1.12em
 \raise.6ex \hbox{$<$} \kern.1em}}
\def\simg{\hbox{\kern.1em \lower.6ex \hbox{$\sim$} \kern-1.12em
 \raise.6ex \hbox{$>$} \kern.1em}}

\newcommand{\beqar}{\begin{eqnarray}}
\newcommand{\eeqar}[1]{\label{#1} \end{eqnarray}}

\begin{document}


\title{Statistical analysis of excitation energies in the
actinide  and rare-earth nuclei}

\author{A.I. Levon}
\email{Email: levon@kinr.kiev.ua}
\author{A.G. Magner}
\email{Email: magner@kinr.kiev.ua}
\author{S.V. Radionov}
\email{Email:sergey.radionov@matfys.lth.se}
\affiliation{\it  Institute for Nuclear Research, 03680 Kyiv, Ukraine}
\bigskip
\date{July, 1st, 2017}
\bigskip

\begin{abstract}

Statistical analysis of
distributions of the collective states
in the actinide and rare-earth nuclei is
performed in terms of the nearest neighbor  spacing
distribution (NNSD).
Several approximations, such as the linear approach
to the level repulsion density and that suggested by Brody to the
NNSDs were applied for the analysis.
We found an intermediate character of the experimental spectra
between the order and the chaos
for a number of the rare-earth and actinide nuclei.
They are more close to the Wigner distribution for
energies limited by 3 MeV,
and to the Poisson distribution
for data including  higher excitation energies and higher
spins. The latter is in agreement with the theoretical calculations.
These features are confirmed by the cumulative distributions, where
the Wigner contribution dominates at smaller spacings
while the Poisson one is more important at larger spacings.
\end{abstract}

\maketitle

\noindent


\section{Introduction}

The microscopic many-body interaction of particles of the
systems, such as heavy  deformed nuclei,
is rather complicated. Therefore, theoretical approaches
to the description of  nuclear excitations
are helpful for understanding
the properties of the collective motion  in
such nuclei. As the simplest approaches, one can mention
 calculations within the
phenomenological interacting-boson model
\cite{Iach87} and a more microscopic quasiparticle-phonon model
\cite{Sol92}. Toward the microscopic picture, other approaches are
described in  Refs.~\cite{Bho,RingSchuck,Migdal}.
However, one can significantly simplify
the realistic many-body problem and enrich its understanding
by using the nuclear models
 which are based on the
statistical properties of the distributions of
discrete levels.

Different statistical methods  have been
proposed to obtain information on the chaoticity versus regularity
in quantum spectra of a nuclear many-body system
\cite{Mehta91,Aberg02,Weiden09,Mitch10}, see also
the well known work by Bohigas, Giannoni and 
Schmit \cite{Bohigas84}.
The short-range fluctuation properties in experimental
spectra can be analyzed in terms of the nearest-neighbor
spacing distribution (NNSD)
statistics. The uncorrelated sequence of energy levels,
originated by a regular dynamics, is described by
the Poisson distribution.
In the case of a completely chaotic dynamics, the
energy intervals between levels
follow mainly the Wigner (Gaussian orthogonal-ensemble)
distribution. An intermediate degree of chaos in
energy spectra is usually obtained through
a comparison of the experimental NNSDs
with well known distributions \cite{Wigner51,Bro73,Berrob84,Izrail88}
based on the fundamental works
\cite{Porter65,Berry81,Brody81,Bohigas84}. This comparison is carried out
\cite{Shrin91,Shrin90,Shrin04,Vidmar07,Dietz17}
by using the least square-fit technique.
The estimated values of parameters of these distributions
shed the light on the statistical situation with considered spectra.
 Berry and Robnik  \cite{Berrob84}
derived the NNSD starting from the microscopic expression for the level density
through the Hamiltonian for
a classical system. The Brody NNSD \cite{Bro73} is
based on the expression for the level repulsion density
that interpolates
between the Poisson and the Wigner
distribution by only one parameter.

For a quantitative measure of the
degree of chaoticity of the many-body dynamics,
the statistical probability
distribution $p(s)$ as function
of  spacings $s$ between the nearest neighboring levels
can be derived within
the general Wigner-Dyson (WD) approach based
on the level repulsion density $g(s)$ (the units will be specified later)
\cite{Porter65,Brody81,Mehta91,Aberg02},
\begin{equation}
p(s) \propto g(s)\;
\exp\left(-\int_0^s g(s')\;\mbox{d}s'\right)\;.
\label{WDgen}
\end{equation}
This approach can be
applied in the random matrix theory, see for instance
Refs.~\cite{Brody81,Weiden09}, and also, for systems with definite
Hamiltonians  \cite{Mehta91,Aberg02}.
In any case, the order in such systems is
approximately associated with the Poisson
dependence of $p(s)$ [Eq.~(\ref{WDgen})] on
the spacing $s$ variable, that is obviously related to a constant
$g(s)$, independent of $s$.
A chaoticity
can be referred, mainly, to  the Wigner distribution,
as clearly follows from
Eq.~(\ref{WDgen}) for $g(s)\propto s$.

For a further study of the order-chaos properties of
nuclear systems, it might be worth
to apply a simple  analytical approximation to
the
WD NNSD (\ref{WDgen}),
keeping the link with a level repulsion density $g(s)$
\cite{Brody81,Mehta91,Aberg02}.
For analysis of the statistical properties
in terms of the Poisson and Wigner distributions,
one can use the linear WD (LWD) approximation to
the level repulsion density  $g(s)$ \cite{BMprc12}.
It is the two-parameter approach; in contrast, e.g., to  the
one-parameter Brody approach. However, the LWD
approximation, as based on a smooth analytical (linear)
function $g(s)$ of $s$,  can be
founded within the WD theory  (see Refs.~\cite{Aberg02,BMprc12}
and Appendix).
Moreover, it gives  a
proper information on the separate Poisson order-like and
Wigner chaos-like contributions.

In the present work, the two different
approaches, such as the LWD approximation to the NNSD (\ref{WDgen})
of the WD  theory,
and the  traditional Brody method
are used for the statistical description of the
collective-excitation energies in
deformed actinide and rare-earth nuclei. This is alternative problem
to that for the nuclear states of another nature; see, e.g.,
Refs.~\cite{Gomez11,Dietz17}.
The statistical properties of the nuclear collective states
are discussed in relation to the degree of
chaoticity in terms of the Poisson and Wigner
distribution contributions. The main purpose is to describe these
excitations in deformed nuclei by using the NNSDs,
in contrast to the states which can be considered as statistically
excited ones in a heated system. In addition, the  cumulative NNSDs
show the statistical properties of these collective states as functions
of the spacing variable in relation to the same limits.

This article is organized in the following way.
Section \ref{sec2:expunfold} is devoted to the description of experimental
data, their completeness and the unfolding
procedure for calculations of the NNSD.
In Section~\ref{sec3:WD},
we present several analytical approximations to the NNSD
within the Wigner-Dyson theory. The NNSD using
the linear approximation to the
level repulsion density, and the Brody  approach
as well as  the  cumulative distribution method (Sec.~\ref{sec3:WD})
are compared with the experimental data in Sec.\ \ref{sec4:disc}.
 Our results are summarized in Sec.~\ref{sec6:concl}.
Some details of our derivations
are given in Appendix.

\section{Experimental NNSD}
\label{sec2:expunfold}

\subsection{Experimental data}\label{sec2-1:data}

To perform statistical analysis of the energy spacings,
one needs the complete and pure sequences of levels.
The completeness means no
missing and no  misassigned  levels
in the desired energy interval of the level sequence.
The problem of missing levels in spectral statistics was considered for
the first time by Bohigas and Pato \cite{Bohigas04} and  reviewed by
Gomez et al. \cite{Gomez11}.
For nuclear physics, the requirement of purity
is that the levels with the same angular-momentum and parity quantum
numbers should be considered in a given interval.
The additional quantum numbers
can be taken into  account
in some cases. Shriner et al. \cite{Shrin91,Shrin90,Shrin04} achieved
these purposes by limiting the energy interval to have
  well-defined, at that time,  spins and parities.
 As a consequence, the number of selected levels
with a particular spin and parity in each nucleus
was usually limited to 5 - 8. As shown in Table \ref{Number_Lev},
much longer sequences  of levels 
\cite{Lev09,Lev13,Lev15,Spi13,Les02,Buc06,Mey06,Bet09,Lev17,Spi17}
are analyzed  in the present study, what
is important for a statistical
accuracy of their fitting procedure.

\begin{figure}
\vskip1mm
\includegraphics[width=7.0cm]{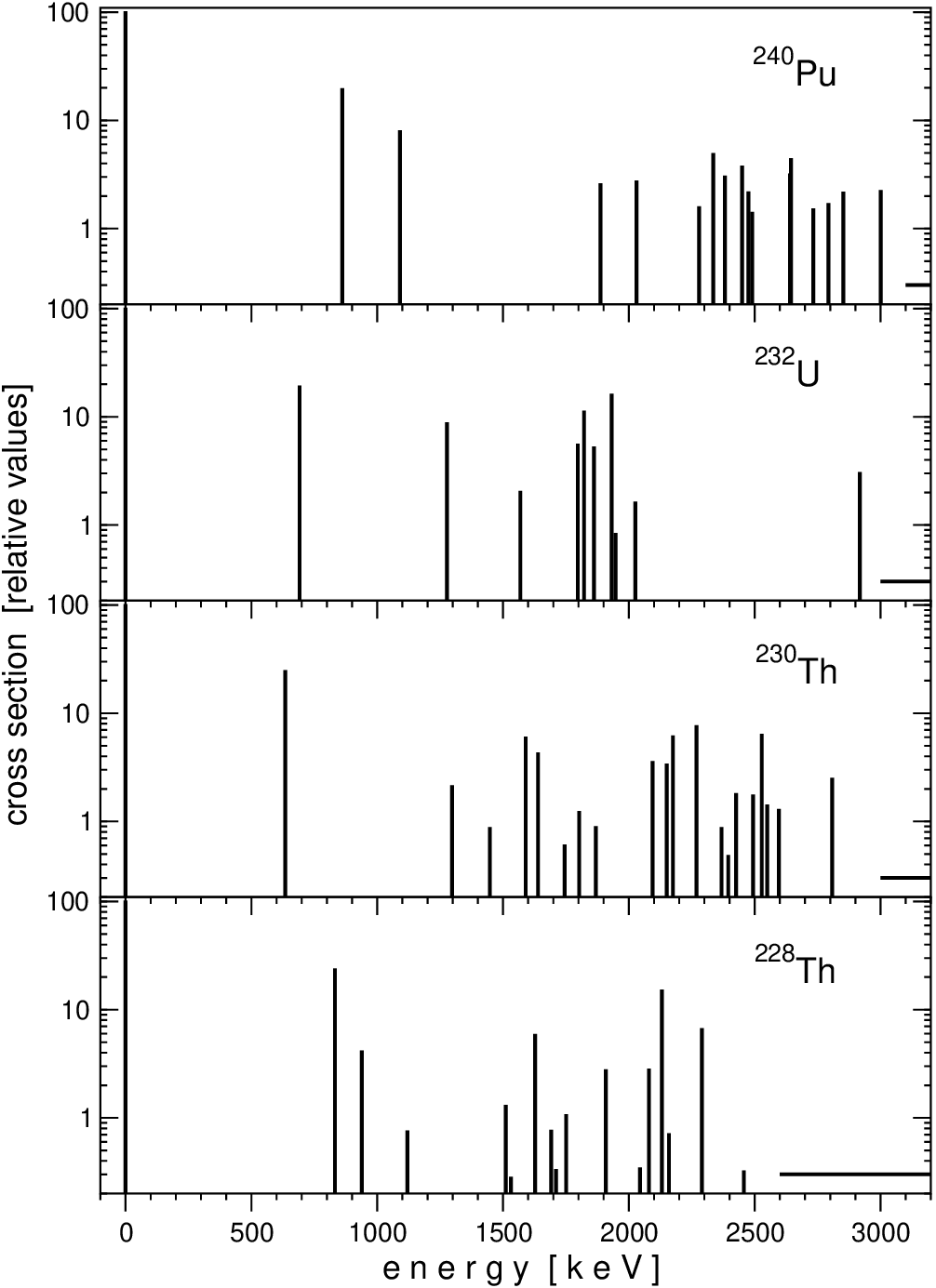}
\vspace{0.6cm}
\vskip-3mm\caption{{\small
The location and
the (p,t) strength of 0$^+$ states  in  $^{228}$Th,  $^{230}$Th,
$^{232}$U and $^{240}$Pu.  Horizontal lines indicate limitations in
the investigation energy.}}
\label{fig1}
\end{figure}

Such sequences are available at  excitations of nuclei in the
two-neutron transfer reactions.
Most of the studies using  these  (p,t) reactions
are devoted to
the investigation of the nature of 0$^+$ states.
In such reactions, one can indeed observe
long sequences
of the collective 0$^+$ states.
The first observation of multiple excitations with the zero
angular-momentum transfer was realized
for the (p,t) reaction
in the odd nucleus $^{229}$Pa \cite{Lev94}.
Such studies were undertaken later by many collaborations, e.g.,
in the deformed even-even actinide
\cite{Lev09,Lev13,Lev15,Spi13}
and  rare-earth
\cite{Les02,Buc06,Mey06,Bet09} nuclei.
Typical spectra of 0$^+$ states are shown in Fig.~\ref{fig1}.
However, the use of the (p,t) reaction is not limited only to the
observation of 0$^+$ states. Long sequences of the states with
higher angular momenta  2$^+$, 4$^+$, and even 6$^+$, along with
the  0$^+$ states, were identified for nuclei in
the actinide  region \cite{Lev09,Lev13,Lev15,Lev17,Spi17}.
They can be used in  the statistical analysis, too.
As emphasized above, the purity of all these sequences of states is
maintained by the fact that all these states are collective by their
nature. This follows from an analysis
within, e.g., the framework of the interacting boson model,
and the quasiparticle phonon model
\cite{LoI05, Buc06, Lev09, Lev13, Lev15}.

\begin{table}
\caption{\label{Number_Lev} Number of levels included in the analysis.}
\begin{ruledtabular}
\begin{tabular}{cccccc}\\
Nuclei & 0$^+$ & 2$^+$ & 4$^+$ & 6$^+$ & Total  \\
\hline\\
$^{228}$Th   & 16   & 32 & 21 & 9   & 78   \\
$^{230}$Th   & 20   & 68 & 46 & 19   & 153   \\
$^{232}$U    & 13   & 46 & 33 & 19  & 111 \\
$^{240}$Pu   & 17   & 37 & 30  & 11 & 95  \\
Rare Earths $\leq$ 3MeV      & 128  &&&&128\\
$^{158}$Gd$^{168}$Er $\leq$ 4MeV  & 58 &&&&58\\
\end{tabular}
\end{ruledtabular}
\end{table}

Excitations of the $0^+$ states have an advantage over those
of other states. Even the weakly excited $0^+$ states,
even in the complicate and dense excitation spectra, are
easily identified via
angular distributions of the cross section in
(p,t) reactions. Shapes of these
angular distributions are mainly independent of specific
structures  of the individual states as well as  of the transfer
configurations. A few levels included
in the analysis are assigned tentatively:   4 of 78 levels
of actinide, and 18 of 128 levels  for  rare-earth
nuclei in the energy region below 3 MeV.
 However, all of 58 levels in the $^{158}$Gd and $^{168}$Er nuclei
 for the interval 0 - 4 MeV are firmly assigned.
Therefore, the  spectra of $0^+$ states measured  in the (p,t) reaction
can be considered as complete ones in energy intervals mentioned
above, that
allows to perform properly the statistical analysis.

Concerning higher spin measurements, one can
outlook the situation
with a sequence of the 2$^+$ levels
in the nucleus $^{230}$Th, as a typical example.
Before the excitation spectra in $^{230}$Th were studied
by means of the (p,t) reaction, the firm  2$^+$ assignments
were known only for 6 levels
and tentative assignments for 21 levels,
mainly, as states with 1 or 2$^+$ spin values, which are denoted by
(1,2$^+$) \cite{Avok93,Acker94}.
In Ref.~\cite{Lev09}, the 66 levels with the spin 2$^+$ were
identified in the (p,t) experiment. The energies of the 5
of 66 levels, within the limit of errors, coincide with those
of (1,2$^+$) known earlier. Moreover, five of previously known
level assignments have been changed \cite{Lev09}  with other quantum numbers.
The remaining  eleven levels with a tentative assignment as
(1,2$^+$) were not observed in the (p,t) reaction
because, probably, they have the spin 1. Such states are not
observed, practically, in the (p,t) reaction. Another reason is that
these states are not collective.
We emphasize once more that a collectivity is
the additional condition for
selections of the level sequences.
Therefore, the 66 levels were analyzed in the present work. Nevertheless,
let us assume that, for completeness,
some of eleven levels should be included
in the sequence. Then, one finds a   
shift of the NNSD to the Poisson distribution, which is additional
to that discussed in Sec.~\ref{sec4:disc}.

In the case of spin  4$^+$,  30\% of
levels are assigned as tentative results.
Their angular distributions exclude the
reliable assignments of 0$^+$ and 2$^+$ spins.
In addition, the (p,t)  cross section for higher spins decreases by almost one
 order of the magnitude. Therefore, for these levels, one can accept the 4$^+$
spin value.
Thus, all the 66 levels with the spin 4$^+$   were included into
our analysis.

Notice that many of 6$^+$ levels are missing
because of a sharp decrease of
the (p,t) cross sections for a so large angular momentum.
 This is particularly true for  the states that correspond to small
values of $s$: Weak peaks in a very complex and dense spectrum  can be hidden
in the tails of stronger neighbors.  As a result, the sequence of
6$^+$ levels occurs to be not complete.
The effect of missing levels in the case of
6$^+$ levels
 is properly discussed in
Sec.~\ref{sec4:disc}.

\subsection{Unfolding procedure}\label{sec2-2:unfold}

To compare properly
the statistical properties of different sequences to each other,
one should convert
any set of the energy levels
into a set of  the
normalized spacing,
that can
be done through  the so-called
unfolding procedure \cite{Boh84}.  In this procedure,
an original set of the level energies  $E_i, i=1,2,...,$ is transformed
into a new set $\varepsilon_i$ as a mapping
\begin{equation}
\varepsilon_i=\widetilde{N}(E_i)\;,
\label{vareps}
\end{equation}
where $\widetilde{N}(E)$
is a  smooth part of the
cumulative level
density, $N(E)=\int_0^E dE'(\mbox{d} N/\mbox{d} E')$,  where
$\mbox{d} N/\mbox{d} E$ is
the level density. The cumulative density $N(E)$
is a staircase function that counts
 the number of states with energies $E$, which are less or equal to
 a given value of $E$.
The decomposition of
this density (or, the level density
itself) into a smooth and fluctuating part is not obvious. Usually, one
can use polynomial fits to  a smooth part.

 In what follows,
the spectra will be analyzed in terms of the
spacings between
the unfolded energy levels (\ref{vareps}),
\begin{equation}
s_i=\varepsilon_{i+1}- \varepsilon_i\;.
\label{si}
\end{equation}
By Taylor expansion of
$\widetilde{N}(E_i)$ in
Eq.~(\ref{vareps})
up to linear terms in $E_{i+1}-E_{i}$, one writes
\begin{equation}
\varepsilon_{i+1}- \varepsilon_i \approx
(E_{i+1}-E_{i})\frac{\mbox{d}\widetilde{N}(E_i)}{\mbox{d} E_i}
=\frac{E_{i+1}-E_{i}}{D_i}\;,
\label{Taylor}
\end{equation}
where $D_i=1/[\mbox{d}\widetilde{N}(E_i)/\mbox{d} E_i]$
is the average level spacing
locally in
 a small
vicinity of $E_i$. For the approximation
(\ref{Taylor}),
one  assumes that the
average dimensionless spacing between the unfolded levels (\ref{vareps}) is
one,
provided that the smooth level density
\begin{figure}
\vskip1mm
\includegraphics[width=7.0cm]{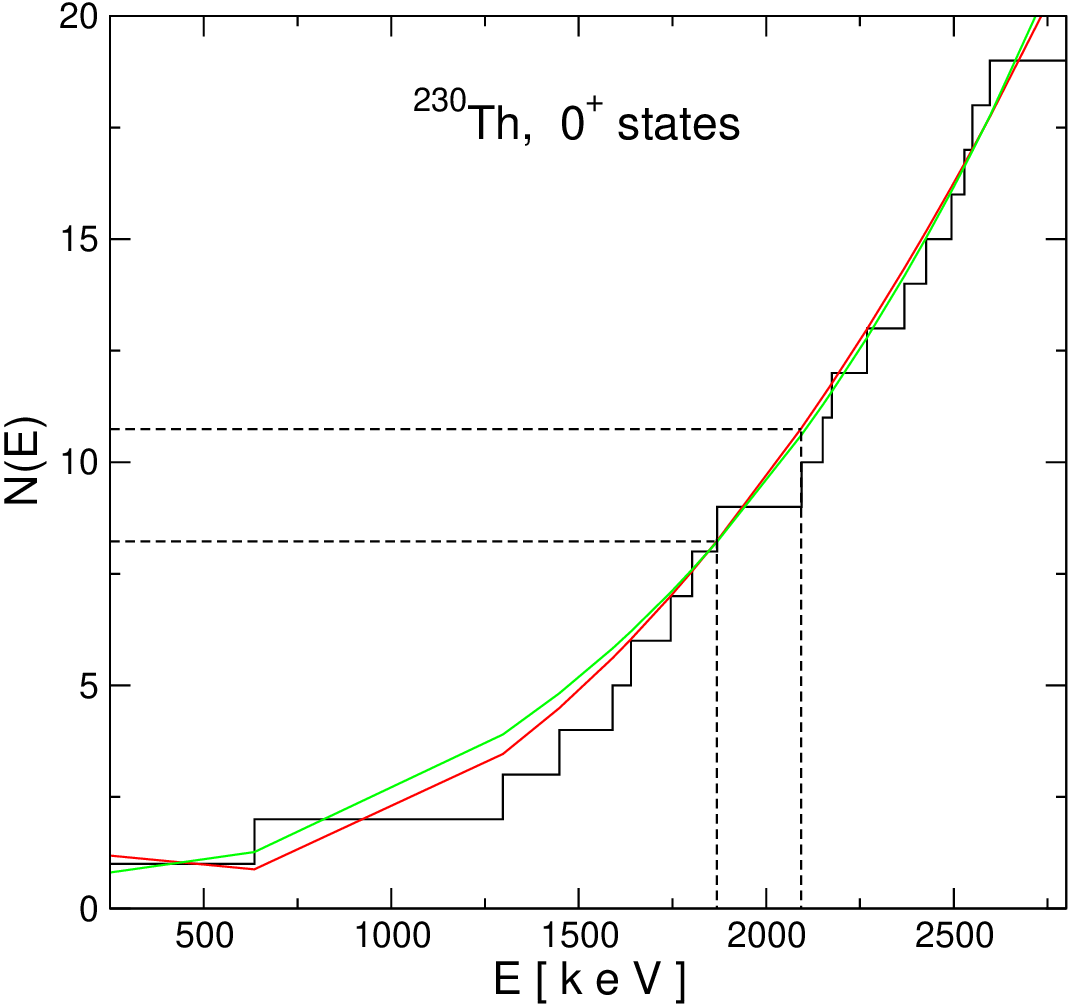}
\vspace{0.7cm}
\vskip-3mm
\caption{{\small Histogram of the cumulative number of states, $N(E)$,
for the 0$^+$ energy spacings in $^{230}$Th  and its fitting by two
smooth polynomials, Eq.~(\ref{polfit}), present  the
method of  an extraction of  the normalized effective
 NNSD  from the
experimental data.}}
\label{fig2}
\end{figure}
$\mbox{d}\widetilde{N}(E_i)/\mbox{d} E_i$ is
a slowly varying
function of the energy $E_i$.

Thus, for each observed level,
the value of the fit function $\widetilde{N}(E)$ can be
used for the generation of a spacing
distribution, as illustrated in Fig.~\ref{fig2}. As one can see,
 a small (large)
energy  spacing
corresponds to
a small (large) spacing
in $\widetilde{N}(E)$, according to the monotonic
mapping [Eqs.~(\ref{si}) and
(\ref{vareps})]. The
distribution $\widetilde{N}(E)$ was then used for
building the final NNSDs.
 Since the
experimental data for a particular sequence
 is statistically limited we compiled the distribution
for all
sequences to get the entire  unified
set of  the nearest  neighbor spacing
distribution  with  relatively a small
sampling interval $\gamma_s$ (see
Sec.~\ref{sec4:disc} and Appendix).

As shown in Fig.~\ref{fig1},
the spectrum of states  is
sparse and widely spread.  Therefore,
the polynomial fitting looks as a difficult problem.
Indeed, using different polynomials to fit the full spectrum of states,
one obtains a different fit at the beginning of the spectrum.
As a result, one finds  somewhat a different NNSD.
It is caused by a specific property of the energy spectra:
The spacing between the  two lowest levels is much larger than
that between all other levels.
In the  statistical analysis,
the first energy interval
contributes to the
NNSD in the region of too large $s$.  Therefore, this level
can be discarded without any completeness violation.
Then, the remaining smooth-state
trend is well reproduced by any low--order polynomial,
and various polynomials
lead to a very close  NNSDs. In the analysis we used
the following polynomials:
\begin{equation}
\widetilde{N} = a^{}_0+a^{}_1E+a^{}_2E^2,~~
\widetilde{N} = a^{}_0+a^{}_1E^2+a^{}_2E^4,
\label{polfit}
\end{equation}
  where $a_i$ are fitting parameters.
Absolute values of $s_i$ for each level, obtained with different polynomials,
 are distinguished somehow.
But the  final result for spacing
distributions  differs only by
the part with a low statistics. Therefore,  the NNSD results
based on both polynomials are stable and compiled well in our
calculations.

Note that another unfolding procedure \cite{Gilbert65}
using an empirical formula
$~\widetilde{N}(E) = \exp[(E-E_0)/T]+N_0,~$
where $T, E_0$, and $N_0$ are the fitting parameters,
was applied \cite{Shrin91} for the statistical
analysis. As was shown \cite{Dietz17} for excited states in  the
spherical nucleus $^{208}$Pb,
both procedures yield
approximately similar results  for NNSDs.
We found that
the polynomial empiric functions
$\widetilde{N}(E)$ [Eq.~(\ref{polfit})]
are more suitable for the statistical analysis of
the collective excitations
in cold deformed nuclei. In addition, we point out  that, 
according to discussions in
Ref.~\cite{Gomez11},
the choice of an unfolding procedure does not influence much  on
the short-range spectral statistics in terms of the NNSDs calculated here.

The NNSD obtained in such a way are normalized to one.
Then, they are fitted by
simple theoretical distributions.

\section{Wigner-Dyson NNSD}
\label{sec3:WD}

\subsection{General ingredients}
\label{sunbec31:geningr}

Following the review
\cite{Brody81},  one  obtains
the  probability $p(s)$
of finding the
spacing $s$ between the nearest neighboring levels,
(Eq.~(\ref{WDgen}) and
Refs.\ \cite{Wigner51,Porter65,Mehta91,Aberg02}),
\begin{equation}
p(s)=\aleph^{-1}~g(s)\;
\exp\left(-\int_0^s g(s')\;\mbox{d}s'\right).
\label{WD}
\end{equation}
As mentioned in Introduction, the key quantity $g(s)$
is the  level repulsion  density, $g(s)=\mbox{d}N/\mbox{d}s$,
where $\mbox{d}N$
is the number of states in the interval $\mbox{d} s$
from $s$ to $s+\mbox{d} s $
(see Appendix). It is convenient
to consider $s$ in units of the average $D$ of distances
between levels, $s=S/D$, where $S$ is the energy spacing, i.e., the
distance between the neighbor levels in  usual energy units.
Thus, $D$ is locally a mean
distance between neighboring levels in  energy units.

Practically,
the normalization factor $\aleph$ of the probability distribution
(\ref{WD}) can be
found with any accuracy for a large
maximal value of $s$, $s_{\rm max}$,
\begin{equation}
\aleph=\int_0^{s_{\rm max}}\mbox{d}s\; g(s)
\;\exp\left(-\int_0^s g(s')\mbox{d}s'\right)\;.
\label{norm}
\end{equation}
This normalization factor
is relatively obtained from
the normalization  condition
at  $s_{\rm max}$ going to $\infty $:
\begin{equation}
\int_0^{s_{\rm max}} p(s)\;\mbox{d}s=1\;.
\label{normcond1}
\end{equation}
Another normalization condition writes
\begin{equation}
\int_0^{s_{\rm max}}  s\; p(s)\;\mbox{d}s= 1\;.
\label{normcond2}
\end{equation}
It is convenient to keep formally the upper integration limit $s_{\rm max}$
as a large
finite number by reasons explained below.
Notice that, also for convenience,
we introduced the
dimensionless  quantities, such as the probability
 distribution $p(s)$, and the
level repulsion density $g(s)$ as functions of
the dimensionless spacing variable $s$
[in Ref.\ \cite{Brody81}, the probability density is denoted by $P(S)$,
where $S=s D$, and
the level repulsion density --
by $r_{10}(S)$].

With the definition of the dimensionless
density $g(s)$, for the uniform case one has $g(s)=1$.
This corresponds, in usual
energy units, to the
energy density $1/D$.
Substituting this constant
level density $g(s)$ into Eq.\ (\ref{WD}), one has the Poisson law
\begin{equation}
p^{}_{\rm P}(s)
=\exp\left(-s\right)\;.
\label{pois}
\end{equation}
The Wigner law follows from the assumption of the
level repulsion density that is proportional to $s$.
In this case, from Eq.~(\ref{WD}) one finds
\begin{equation}
p^{}_{\rm W}(s)
=\left(\pi s/2\right)\;\exp\left(-\pi s^2/4\right)\;.
\label{wig}
\end{equation}
Both distributions are normalized to one for a large
maximal value of $s$ in order to satisfy
Eqs.\ (\ref{normcond1}) and (\ref{normcond2}) at large $s_{\rm max}$
and, precisely, at $s_{\rm max} \rightarrow \infty$.

The density $g(s)$ in fact is not a constant or
simply proportional to $s$.
A simple distribution based on the two-parameter linear
approximation to the level repulsion
density $g(s)$, that bridges the both Poisson (\ref{pois})
and Wigner (\ref{wig}) limits, will be considered first in the next section.

\subsection{A linear level-repulsion density approximation}
\label{subsec32:LWD}

Keeping a link with the analytical properties of the
level repulsion density $g(s)$ (Appendix), it is
convenient to define the probability
$p(s)$ [Eq.\ (\ref{WD})] for a general smooth
density $g(s)$ as a polynomial of not too a large power. As shown
in Appendix, it is important to care of this density smoothness. For the
simplest statistical analysis in terms of the Poisson- and Wigner-like
distribution contributions, one can use the linear approximation of
$g(s)$
in terms of the two free parameters
$a$  and $b$,
\begin{equation}
g(s)=a + b s\;.
\label{denlin}
\end{equation}
Substituting Eq.\ (\ref{denlin}) into the general Wigner-Dyson
formula (\ref{WD})  and using the normalization condition (\ref{normcond1})
with large but a finite upper limit $s_{\rm max}$  (larger than the
experimental data),
one obtains explicitly the analytically simple
LWD approximation \cite{BMprc12}:
\begin{equation}
p(s)=\frac{1+b s/a}{\aleph_0 + b\;\aleph_1/a}\;
\exp\left(-\frac{b}{2} s^2 - a s\right)\;,
\label{pslinPW}
\end{equation}
where
\begin{eqnarray}
\aleph_0&=&
\int_0^c \mbox{d}s \;
\exp\left(-\frac{b}{2} s^2 - a s\right)\nonumber\\
&=&
\sqrt{\frac{\pi}{2b}}\;\exp\left(\frac{a^2}{2 b^2}\right)\;
\left[\mbox{erf}\left(\frac{a+b c}{\sqrt{2b}}\right)-
\mbox{erf}\left(\frac{a}{\sqrt{2b}}\right)\right],
\nonumber\\
\aleph_1&=&
\int_0^c \mbox{d} s\; s\;\exp\left(-\frac{b}{2} s^2 -
a s\right)\nonumber\\
&=&
\frac{1}{b}\left[1-\exp\left(-\frac{b}{2} c^2 -
a c\right)- a\;\aleph_0\right]\;,
\label{pslin}
\end{eqnarray}
 with
$c=s_{\rm max}=S_{\rm max}/D$ being the maximal value of $s$.
Then, we should check the second normalization condition
(\ref{normcond2}) by choosing the parameter $c$ larger than
all of the experimental NNSD spacings $s$.
In practice, it is convenient to perform
the three-parameter fitting over parameters $a$, $b$, and $c$ to
the experimental NNSD provided that the normalization
condition (\ref{normcond2}) is satisfied,
and then, check that
$c$ is sufficiently large
with a good accuracy.
In the limit $c \rightarrow \infty$, one has simply
\begin{equation}
\aleph_0 \rightarrow \sqrt{\frac{\pi}{2 b}}\;
\exp\left[\frac{a^2}{2b^2}\right],\quad
\aleph_1 \rightarrow \frac{1}{b}\left[1-a\sqrt{\frac{\pi}{2b}}\right].
\label{alephas}
\end{equation}
Taking the limits $a \rightarrow 1$, $b\rightarrow 0$
and $a \rightarrow 0$, $b\rightarrow \pi/2$ in Eq.~(\ref{pslinPW}),
one simply arrives relatively at
the standard Poisson $g^{}_{P}(s)$, Eq.\ (\ref{pois}), and
Wigner  $g^{}_{W}(s)$, Eq.\ (\ref{wig}), distributions. In this way,
a linear approximation
(\ref{denlin}) unifies analytically these two limit cases through  a smooth
level-repulsion density $g(s)$.
Its parameters
$a$ and $b$ in  Eq.\ (\ref{denlin})   (after their normalization to one
for convenience)
measure the probability to have separately
the Poisson and Wigner distribution contributions.

\subsection{The Brody distribution}
\label{subsec33:brody}

 The Brody distribution can be derived analytically
from Eq.~(\ref{WD}) by
assuming
the following expression for the level repulsion density:
\begin{equation}
g(s)=\alpha\; s^q\;.
\label{gBrody}
\end{equation}
 With the normalization
condition (\ref{normcond1}),
another condition (\ref{normcond2}) is satisfied identically.
Finally, one finds \cite{Bro73,Brody81,Gomez11}
\begin{equation}
p^{}_{\rm B}(s)=
\alpha(q)\; s^q \;
\exp\left[-\frac{\alpha(q)}{q+1}\; s^{q+1}\right]\;,
\label{PBrody}
\end{equation}
where
\begin{equation}
\alpha=\left(1+q\right)\;
\left[\Gamma\left(\frac{q+2}{q+1}\right)\right]^{q+1}\;.
\label{alphaeq}
\end{equation}
Here, $\Gamma(x)$ is the standard Gamma function, and $q$ is a free parameter.
The values $q=0$ ($s>0$) and $q=1$ ($s \geq 0$) in Eq.~(\ref{PBrody})
correspond to the same Poisson [Eq.~(\ref{pois})]
and Wigner [Eq.~(\ref{wig})] distributions.

The only one parameter is  an
advantage of the
popular distribution $g^{}_{B}(s)$  [Eq.~(\ref{PBrody})],
suggested by Brody,
over the approximation
(\ref{pslinPW}) based on the linear
level-repulsion density $g(s)$.
As compared to the Brody approach, the two-parameter LWD
approximation (\ref{pslinPW}) is, to some extent,
more general and better founded
within the WD analysis in terms of
the ordered Poisson and chaotized Wigner distributions.
 As a linear approximation (\ref{pslinPW})
for $g(s)$, it has
a more clear meaning of the intermediate values of the
parameters, found from the least-square fitting to the
experimental  NNSD.
In this way, one obtains
the separate
Poisson and Wigner distribution  contributions.
This is in addition
to the
Brody distribution (\ref{PBrody})
based on the
power density (\ref{gBrody}).
Such a density [Eq.~(\ref{gBrody})]
does not satisfy
a smoothness property of the
level repulsion
densities $g(s)$, in spite of using it
in derivations of the NNSD within
the WD theory
(see Appendix). However, as shown below the results obtained by the
LWD and the Brody approach are largely well agreed to each other.

Thus, the probability density (\ref{pslinPW}) is a  simple
analytical continuation
 from
the Poisson  $g^{}_{P}(s)$ to Wigner  $g^{}_{W}(s)$ limit distributions
through  a smooth linear
  level-repulsion density $g(s)$.
 For a comparison and completeness,
the statistical analysis
of the experimentally obtained
excitation-energy distributions are performed below
within both the LWD and
Brody
approximations.

\subsection{Cumulative NNSD}
\label{subsec34:calcumdist}

To complement  our
NNSD analysis of nuclear spectra,
one can use the cumulative NNSDs. 
The cumulative NNSD
is used as an alternative method to study the statistical properties
of the experimental cumulative NNSD
depending on the spacing variable $s$, in addition to the NNSD
\cite{Gomez11,Shrin91}.
In this subsection,  we restrict ourselves to
the additional information
about the nuclear level statistics, depending on
the spacing variable $s$,
from the cumulative spacing distributions to support our NNSD results.
A more proper quantitative study of these
statistical properties of nuclear excitations
will be in forthcoming work.

Let us consider the cumulative nearest-neighbor spacing
distribution,
\begin{equation}
F(s)=\int_0^s p(s')\mbox{d}s'.
\label{Fs}
\end{equation}
This integral distribution is
the probability to
find the
spacing $s'$ between
the two-neighbor
energy levels smaller or equal to a given value of $s$.
For the cumulative Poisson distribution,
one can explicitly obtain from
Eqs.~(\ref{Fs}) and (\ref{pois})
\begin{equation}
F_{\rm P}(s)=1-\exp\left(-s\right)\;.
\label{FP}
\end{equation}
For the  corresponding
Wigner distribution limit of $F(s)$, one finds
\begin{equation}
F_{\rm W}(s)=1-\exp\left(-\pi s^2/4\right)\;.
\label{FW}
\end{equation}
Analytical expressions for the cumulative distributions
obtained easy from the
LWD [Eq.~(\ref{pslinPW})] and the Brody [Eq.~(\ref{PBrody})]
approach to the
probability density will be worked out
later.

\vspace{-0.5cm}
\begin{figure*}
\includegraphics[width=12.5cm]{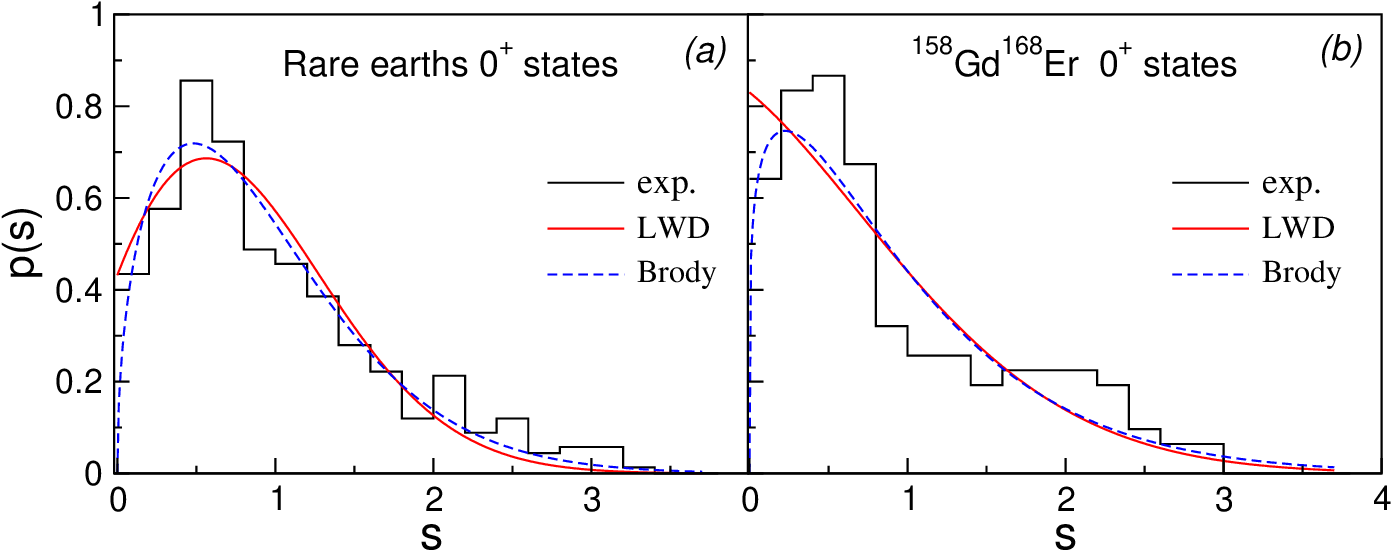}

\vspace{0.7cm}
\caption{{\small
Nearest neighbor spacing distributions
$p(s)$ as functions of a
dimensionless spacing variable $s$ for 0$^+$ states in the rare-earth nuclei
and fits by
the LWD approximation (red solid lines) and
the
Brody approach  (blue dashed lines);
 (a): for a number of the rare-earth nuclei
up to the energy 3 MeV (see
the text);
(b): for the $^{158}$Gd and $^{168}$Er nuclei up to
about 4 MeV.
A sampling interval of $\gamma_s=0.2$
was used. }}
\label{fig3}
\end{figure*}

\begin{figure*}
\vspace{1.5cm}
\includegraphics[width=12.0cm]{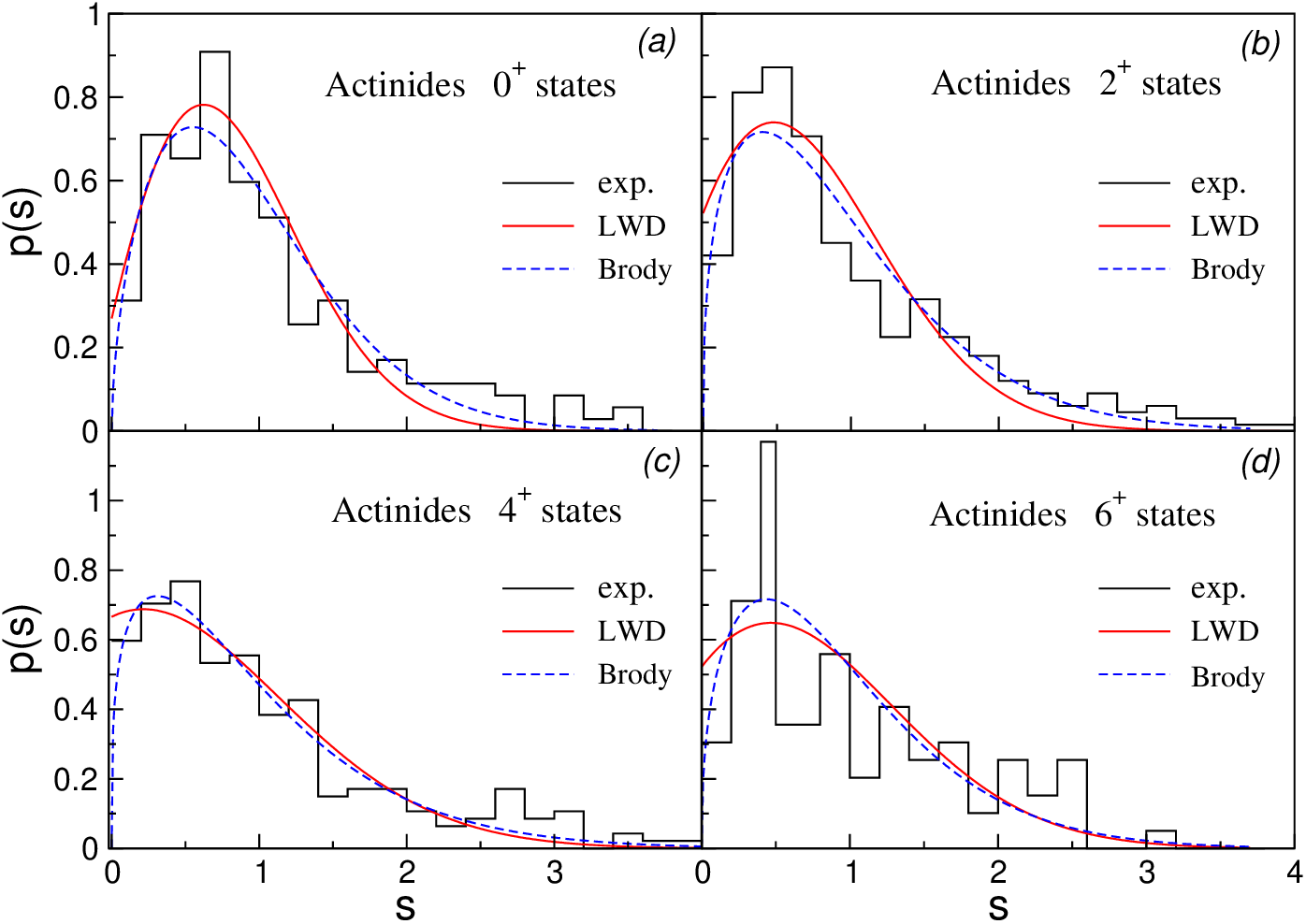}

\vspace{0.8cm}
\caption{{\small
The same as in Fig.~\ref{fig3} but
for different states in the actinide nuclei;
(a-d):
for 0$^+$, 2$^+$, 4$^+$, and
6$^+$ states, respectively.
The same fits
by the LWD
(red solid lines) and
the Brody (blue dashed lines) approach as in Fig.~\ref{fig3} are shown.
}}
\label{fig4}
\end{figure*}

\vspace{-2.5cm}
\begin{figure*}
\includegraphics[width=14cm]{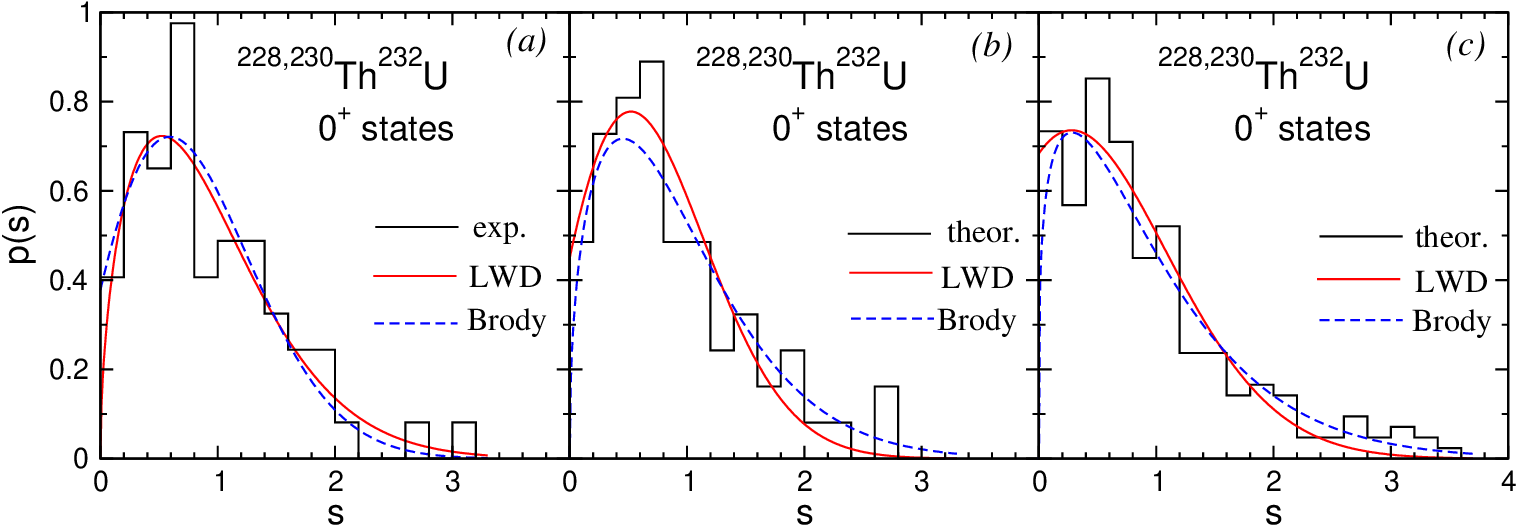}

\vspace{1.2cm}
\caption{{\small
Comparison of the NNSD between the experimental
data $(a)$ and the theoretical quasiparticle-phonon model results $(b)$
in the energy interval up to 3 MeV  in
the $^{228,230}$Th and $^{232}$U actinide nuclei,
and  those $(c)$ - up to 4 MeV.
Other notations are the same as in
Figs.~\ref{fig3} and \ref{fig4}.
}}
\label{fig5}
\end{figure*}

\vspace{3.0cm}
\section{Discussions of the results}
\label{sec4:disc}

Experimental nearest-neighbor
spacing distributions fitted by the LWD approximation
[Eq.~(\ref{pslinPW})]
and the
Brody approach [Eq.~(\ref{PBrody})]
are presented  in Figs.~\ref{fig3} -- \ref{fig5}.
Parameters of fittings are given in
Table \ref{table2}.
The sampling interval $\gamma_s=0.2$,
used for building
the experimental NNSD (Sec.~\ref{sec2:expunfold}),
is taken from the condition of the
stable smoothed NNSD values
without sharp jumps between the neighbor data.
This is similar to the so-called plateau condition in the
smoothing procedure for calculations of the averaged
level density  \cite{strut,FH}.
The plateau condition means the independence of
the averaging parameters.
As follows from  Table \ref{table2},
the normalization condition (\ref{normcond2})
 is satisfied in our calculations with good accuracy at $c\simg 10$,
that is significantly larger than
any of energy intervals found from
experimental data.

\vspace{0.2cm}
\begin{table*}[pt]
\begin{center}
\begin{tabular}{|c|c|c|c|c|c|c|c|c|c|c|}
\hline
Figs. & Nuclei & States & a (\%)&
b (\%)& $\int\! sp(s)\mbox{d} s$& Accuracy & q   & Accuracy\\
\hline
3a &Rare earths & 0$^+$ ~exp.& 0.43 (39) & 0.69 (61)
& 1.04 & 8.1 &  0.48 & 6.7 \\
3b &$^{158}$Gd $^{168}$Er & 0$^+$  ~exp.&0.83 (76) & 0.26 (24) &  0.96
& 11.3 &0.20 & 10.0 \\
\hline
4a &Actinides & 0$^+$  ~exp.& 0.27 (21) & 1.01 (79)
& 1.02 &9.2& 0.58 & 8.9 \\
4b & & 2$^+$  ~exp.& 0.52 (41) & 0.75 (59)
& 0.95 &10.2& 0.38 & 8.2 \\
4c & & 4$^+$  ~exp.& 0.67 (62) & 0.41 (38)
& 1.00  &8.5& 0.28 & 7.3 \\
4d && 6$^+$  ~exp.& 0.52 (50) & 0.52 (50)
&  1.06 &14.9& 0.42 & 13.7 \\
\hline
5a & $^{228,230}$Th, $^{232}$U & 0$^+$  ~exp.
& 0.38 (32) & 0.80 (68)
& 1.03 & 10.5 &  0.54 & 10.1 \\
5b &  &~ 0$^+$ ~theor.
&0.45 (33) & 0.91 (67) &  0.94
& 9.7 &0.44 & 9.5 \\
5c &  & ~~0$^+$ ~theor.
& 0.68 (56) & 0.54 (44)
& 0.93 &8.7& 0.25 & 8.9 \\
\hline
\end{tabular}

\vspace{0.2cm}
\caption{{\small
Parameters $a$ and $b$ of the LWD and $q$ of the Brody
approximation for the collective excited states
in several nuclei.
The 1st column refers to the corresponding figures \ref{fig3}-\ref{fig5}.
The Poisson and Wigner contributions are given
also as  $a$ and $b$ normalized
to 100\% in circle brackets in 4th and 5th columns, respectively.
The normalization integral of Eq.~(\ref{normcond2}) at these
$a$, $b$ and $c\simg 10$ is given too in 6th column.
The accuracy ($\chi^2$)
of the  least-square fitting (in percents)
are shown, respectively, in the 7th and 9th columns for the LWD and
Brody calculations.
}}
\label{table2}
\end{center}
\end{table*}

To build the NNSD for the
 rare-earth nuclei (Sec.~\ref{sec2:expunfold}),
the experimental 0$^+$ state energies
limited by the 3 MeV excitation
are used for $^{158}$Gd  \cite{Les02}, $^{168}$Er  \cite{Buc06},
$^{152,154}$Gd, $^{162}$Dy, $^{168}$Er, $^{176}$Hf, $^{180,184}$W,
 and $^{190}$Os  \cite{Mey06}, $^{170}$Yb  \cite{Bet09} nuclei
[Fig.~\ref{fig3}(a)].
The experimental NNSD for nuclei $^{158}$Gd \cite{Buc06}
and $^{168}$Er \cite{Lev17}
[Fig.~\ref{fig3}(b)] is a special case since only for
these two nuclei the measurements were carried out for larger
excitation energies up to 4.2 MeV.
The results of fitting are following. The
 rare-earth nuclear spectrum is described
by 39\% of the Poisson- and 61\%
of the Wigner-distribution  contribution.
They correspond  approximately to the parameter q=0.48
in the Brody approach.
Simultaneously, for the $^{158}$Gd and $^{168}$Er couple,
these parameters are  given by
76\% and 24\%,  respectively, in the LWD approximation.
 This can be tentatively related to the value q=0.20
for the Brody distribution.
This means that the experimental 0$^+$
spectra in the energy interval 0--3 MeV are intermediate
between an  order and a  little more pronounced
chaos structure, while the ordered nature is  dominant
for the experimental
spectra in the energy interval about 0−-4 MeV.

As was pointed above (Sec.~\ref{sec2-1:data}), the experimental NNSDs for
actinide nuclei are
available also
for 2$^+$, 4$^+$ and 6$^+$ collective states, along with
0$^+$ excitations.
Long sequences of 2$^+$, 4$^+$, and even 6$^+$ states, as well as 0$^+$
states,  all identified in
$^{228}$Th \cite{Lev13}, $^{230}$Th \cite{Lev09},
$^{232}$U \cite{Lev15} and $^{242}$Pu \cite{Spi13,Spi17}, are used
in  our analysis. As  seen from
Table \ref{table2}, the picture is similar to the rare-earth  behavior.
All spectra in the same energy interval 0--3 MeV demonstrate
 an intermediate structure between an order
and a chaos structure with varying dominance of the Wigner to the
Poisson contribution for increasing the angular momentum up to 4$^+$.
If for 0$^+$ states
the Wigner contribution dominates with 79 \% ($q = 0.58$), for the states
with higher angular  momenta, namely, for the
2$^+$ and 4$^+$ states, the Wigner contribution is somehow  lowering.
For example, one can conclude that the
fluctuation properties
for the 2$^+$ states is closer to the Poisson distribution
than those for the 0$^+$ states. See Sec.~\ref{sec2-1:data} for discussions
of the level sequence completeness.
For the 4$^+$
states the Poisson contribution  becomes  even
dominating with  62 \% ($q=0.28$).

The results for 6$^+$ states seem to be different from
this trend. The 6$^+$ NNSDs are also more close to
the Possion distribution as compared to  0$^{+}$ ones while the opposite
tendency takes place with respect to the 4$^{+}$ states:
The Poisson contribution  for  6$^+$ states
is found less
than that for 4$^+$ states.
As  was already pointed out in Sec.~\ref{sec2-1:data},
the sequence for the 6$^+$ levels can not be
apparently reliably
completed because of a sharp drop in the cross-section value
 with increasing the angular momentum. This can be a reason of missing levels
and, therefore, of their NNSD deflection from a general trend
(see Sec.~\ref{sec2-1:data} for more details).
A further progress in experimental studies of such high-spin
collective states can be apparently helpful to clarify more the
6$^+$ statistical
properties.

An increase of the Poisson distribution contribution with
the nuclear spin value can be
considered,  to some extent,
as that with a growth of the energy:
The array of the states with
higher angular momenta is shifted to  larger energy excitations, too,
as compared to the 0$^+$ case.
This behavior of the statistical distribution would look
strange  when accepting that  increasing
the excitation energy means an increase
of the temperature,  or of the thermodynamic entropy. Such an
entropy production could be interpreted as  a growing chaos.
This would mean that
the Wigner distribution contribution should be  greater
for higher energies. However, as it was
emphasized above, one can conclude
about  the collective nature  of
states  excited  in the (p,t) reaction,
see  Refs.~\cite{LoI05, Lev09}.
Collective excitations under the consideration in deformed nuclei
can not be associated with an increase
of  temperature. 
Our results certainly different from those
obtained \cite{Dietz17} for
the complete sets of noncollective
states in nucleus $^{208}$Pb.
For such states, the chaoticity dominates both in the experimental and
spectrum and that calculated
within the shell model. 
Our calculations
are in accordance with the results of Shriner at al. \cite{Shrin91}
and discussions in Ref.~\cite{Gomez11}.
In any case,
our statistical analysis
provides another  view on developments of a more
microscopic model for the theoretical calculations of
the collective modes.

\begin{figure*}
\includegraphics[width=12.0cm]{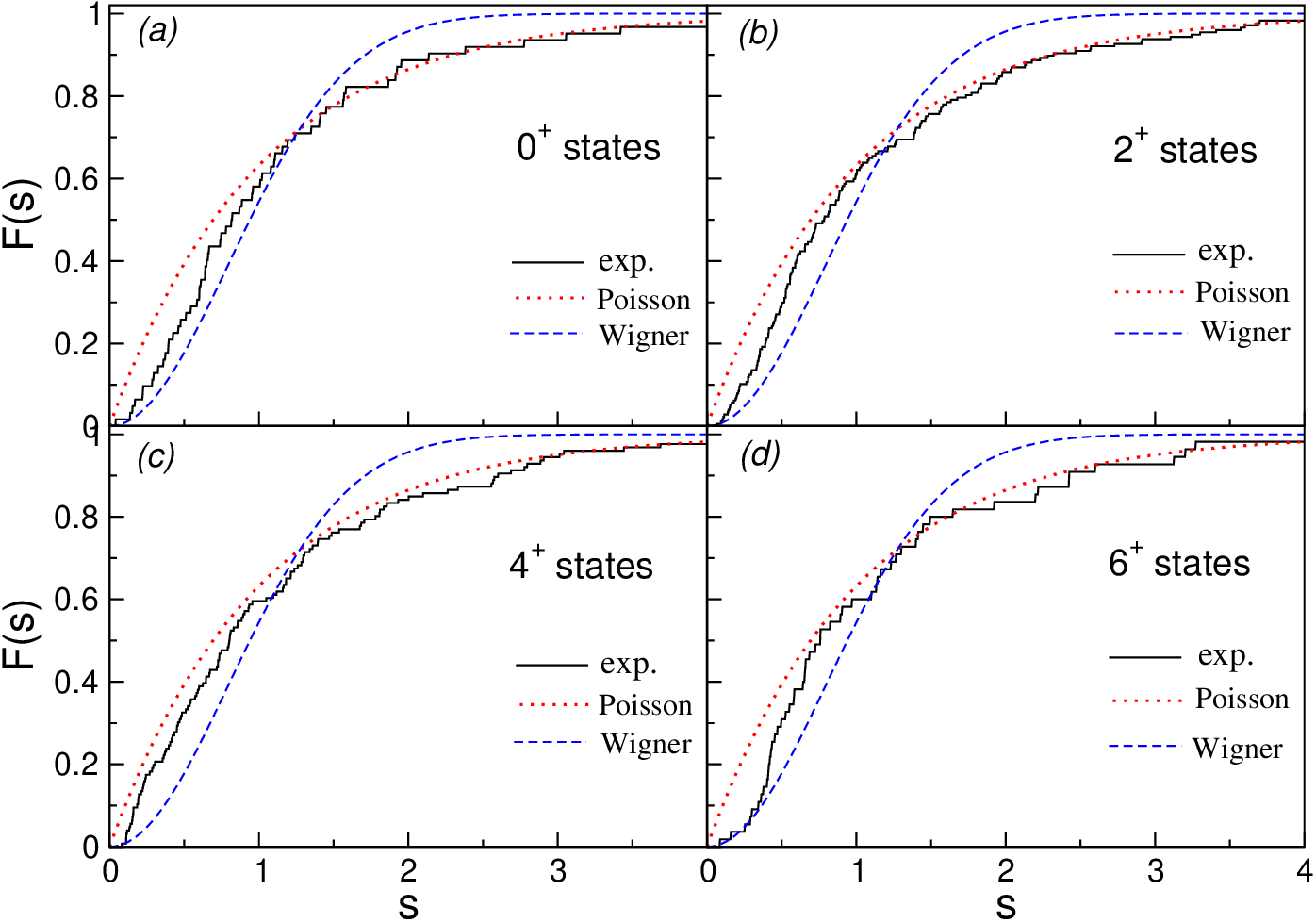}
\vspace{0.7cm}
\caption{{\small Histograms of the cumulative nearest--neighbor
spacing  distributions (\ref{Fs}) for the $0^+$ (a),
$2^+$ (b), $4^+$ (c) and $6^+$ (d) states in the
actinide nuclei discussed
in  Fig.~\ref{fig4}.
Dotted and dashed lines
are the Poisson distribution
[Eq.~(\ref{FP})] and the Wigner distribution
[Eq.~(\ref{FW})] limit of
these cumulative
distributions (\ref{Fs}), respectively. }
}
\label{fig6}
\end{figure*}

Following the ideas of
Refs.~\cite{Shrin91,Dietz17} we tested the validity
and completeness of the level sequences by comparing the
experimental data with the collective state spectra
calculated within
the quasiparticle phonon model  \cite{LoI05,Buc06}. Fig. \ref{fig5}
presents the distributions for 0$^+$ states
in three actinides $^{228,230}$Th
and $^{232}$U. The experimental NNSD
in the region of 0-3 MeV (a) are compared with the two
theoretical distributions.
One of them is given in the same energy region (b)
and another distribution  -- in the extended energy
interval 0 - 4 MeV (c). The parameters for
the distributions shown on the panels (a) and (b)
are approximately the same within the error limit accuracy.
This  agreement between the experimental and theoretical results  confirms
the collective nature of the 0$^+$ states and, finally,
the completeness of the level sequences. At the same time, the theoretical
distribution for the energy interval 0-4 MeV [Fig.~\ref{fig5}(c)]
is shifted to the Poisson law
as compared to the  experimental and theoretical
distributions in the interval 0-3 MeV. It
is in agreement with the results obtained for the
$^{158}$Gd and $^{168}$Er nuclei
(see Fig.~ \ref{fig3}(b)).

Fig.~\ref{fig6} shows the
cumulative distributions $F(s)$ [Eq.~(\ref{Fs})]
for the $0^+$ (a), $2^+$ (b), $4^+$ (c), and $6^+$ (d)
states excited in
the same actinide nuclei as in Fig.~\ref{fig4}.
The dotted lines represent
the corresponding cumulative
Poisson distribution [Eq.~(\ref{FP})].
All dashed lines in Fig.~\ref{fig6} show
the Wigner distribution limit of $F(s)$ [Eq.~(\ref{FW})].
As seen from this Figure,
for all $0^+$, $2^+$, $4^+$ and $6^+$ states
the Wigner cumulative distribution
(\ref{FW}) well reproduces
the behavior of empirical distributions
$F(s)$ [Eq.~(\ref{Fs})] at small and intermediate spacings $s$. On the
other hand, at large spacings, $F(s)$ approaches basically the
Poisson cumulative-NNSD limit (\ref{FP}). Such a
peculiarity of a cumulative distribution
implies a chaotic arrangements of
close-lying levels and regular ones
of the significantly separated
levels. This is in agreement with our results for the NNSD plotted
in Figs.~\ref{fig3}-\ref{fig5}. As in the case of
using the NNSD (see Fig.~\ref{fig4}), the cumulative distribution
analysis of Fig.~\ref{fig6} shows that the relative
Poisson contribution (\ref{FP})
grows with the increase
of the spin of nuclear states.

\section{Conclusions}
\label{sec6:concl}

We provide the statistical analysis of
the collective excitations with several
spins: 0$^+$ in a number of the rare-earth nuclei; and 0$^+$, 2$^+$,
4$^+$, and 6$^+$ in a few actinide nuclei
by using the simple approximations to the
Wigner-Dyson probability distribution. These approximations to the
nearest neighbor spacing
distribution are based
on different properties of the level repulsion density. For the linear
approximation to this density, one obtains a clear information on the
quantitative measure of the Poisson order and Wigner chaos contributions
in the experimental data, separately, in contrast to the heuristic
Brody approach. However, one finds in
our calculations
that the Brody formula [Eq.~(\ref{PBrody})]
agrees largely well with
the LWD probability-distribution results [Eq.~(\ref{pslinPW})].

We found the intermediate structure
between the Poisson and Wigner statistical peculiarities of the
experimental spectra by evaluating their separate contributions.
The NNSD for a smaller
excitation-energy region can be described better by the
Wigner distribution.
The NNSD for an extended interval of the collective
excitations, including higher energies,
becomes more close to the Poisson distribution. Also, one finds
that the Wigner contribution dominates in the NNSD
for 0$^+$ states and the Poisson contribution is larger
with increasing the angular momentum.
This looks in line of the adiabatic picture for different
collective-excitation
modes in deformed nuclei.

The experimental NNSDs are in agreement with the theoretical calculations
for the same energy interval within the quasiparticle phonon model, that
confirms the completeness of the used spectra. The comparison of these
 results with the theoretical ones for larger energy interval supports the same
conclusion about a shift from the Wigner to the Poisson
contribution dominance.
As emphasized in Ref.~\cite{Gomez11},
for the collective states in deformed nuclei the statistical distributions
are closer to
the Poisson distribution, and in other cases the situation is intermediate
(see also Ref.~\cite{Shrin91}). This picture looks in agreement with our
statistical results for the collective states.

With the help of the cumulative distributions, for
the $0^+$, $2^+$, $4^+$ and 6$^+$ states in actinide nuclei
we show that the chaotic cumulative Wigner limit
well reproduces the behavior of
empirical cumulative distributions
$F(s)$ at small and intermediate spacings $s$.
At larger spacings they approach
the regular Poisson cumulative-distribution
limit. In line of
the nearest-neighbor spacing distribution calculations, the
cumulative distribution analysis
shows also that the relative
Poisson contribution
grows with the increase
of the spin of nuclear states.

As
perspectives, we are also going to study more the Wigner-Dyson
probability-density approach within simple approximations
and apply them more systematically
to learn the statistical properties of experimental
data. In this way, it will be worth to calculate
the nearest neighbor spacing distributions for a nonlinear
level-repulsion density to describe other statistically
observable spectra of the collective nature
beyond the Wigner and Poisson contributions.
We are going also to understand the influence of the symmetry breaking
phenomena on these distributions of the collective states
in deformed nuclei.

\section{Acknowledgements}

We are grateful to M.~Spieker for providing us with experimental 
results before publication, also to
K.~Arita, S.~Aberg, J.~Blocki, V.~M.~Kolomietz,
S.~Mizutori, and K.~Matsuyanagi
for many helpful discussions.
One of us (A.G.M.) is also very grateful for kind hospitality during his
working visits of Physical Department of the Nagoya Institute of Technology,
also the Japanese Society of Promotion of Sciences for financial support,
Grant No. S-14130.

\setcounter{equation}{0}
\renewcommand{\theequation}{A\arabic{equation}}

\begin{center}
\textbf{Appendix: The derivation of the NNSD}
\label{appA}
\end{center}

We introduce first the  level
repulsion density, $g(s)$, as the number of the
levels  $\mbox{d} N$ in the dimensionless energy interval
$[e+s,e+s+\mbox{d} S]$,
divided by the energy interval length $\mbox{d} s$,
$g(s)=\mbox{d} N/\mbox{d} s$ \cite{Brody81}.
With the help of this
quantity, one can derive the
NNSD $p(s)$ as the probability density that is
a function of the spacing $s$ between the nearest neighboring levels
in dimensionless units $e=E/D$ and $s=S/D$, where $D$
is locally the averaged distance
between neighbor levels.
Specifying $p(s)$
to the problem with the known
spectra of the many-body (or single-particle) Hamiltonian, one can
split relatively a small energy interval $\Delta e$
under the investigation into many small
(equivalent for simplicity) parts
$\gamma_s \ll \Delta e$. Each of
$\gamma_s$, nevertheless, contain many energy levels.
Then, we find the number of the levels
which occur inside of the
relatively small interval
$\gamma_s$.
Normalizing these numbers by the total number of the levels
inside the total energy interval $\Delta e$, one obtains the distribution
which we shall call as the probability
density $p(s)$.

Notice that the result of this calculation depends on the
spacing length of the selected  $\gamma_s$.
In our calculations, we select
$\gamma_s$ by the condition of a sufficient smoothness of the
distribution $p(s)$. We have to study $p(s)$ as a function of
$\gamma_s$ at a given $s$ for several values of the parameters of
this distribution
to find a so called ``plateau'' in $\gamma_s$, i.e., a
region of $\gamma_s$ values where $p(s)$
can be approximately considered as a constant independent of $\gamma_s$
and above mentioned parameters (see Refs.\ \cite{strut,FH}).
Such a procedure is often used for the
statistical treatment
of the experimentally obtained spectrum with the fixed quantum numbers
as the angular momentum, parity and so on \cite{Brody81}
(Sec.~\ref{sec2:expunfold}).

Following mainly Ref.\ \cite{Aberg02}, let us calculate first the
intermediate quantity
$f(s)$ as the probability that there is no an energy level
in the energy interval $[e,e+s]$. According to
a general definition of the level repulsion density
mentioned above,  $g(s) \mbox{d} s$, can be considered as the probability
that there is one energy level in the interval
$[e+s, e+s+\mbox{d} s]$. Then, one has
\begin{equation}
f(s+\mbox{d} s)=f(s)\left(1-g(s) \mbox{d} s\right)\;.
\label{f}
\end{equation}
Assuming that $f(s)$ is a smooth function of $s$, one can expand
   $f(s+\mbox{d} s)$ with respect to $\mbox{d} s$. Thus, the relationship
(\ref{f}) leads to the differential equation for $f(s)$,
\begin{equation}
\mbox{d} f= -g(s) \mbox{d} s f(s).
\label{diff}
\end{equation}
Solving this equation, one gets
\begin{equation}
f(s)=C \exp\left(-\int_0^s g(s^\prime) \mbox{d} s^\prime\right)\;,
\label{fsol}
\end{equation}
where $C$ is arbitrary unknown constant.
Note that the assumption that $f(s)$ is a smooth function
of $s$ can be satisfied if $g(s)$ is also
a smooth function of $s$, i.e., the density $g(s)$ can be approximated
by a polynomial in powers of $s$ of not too a high power. Notice also
that a constant density, $g(s)=a$, and linear, $g(s)=b s$, functions of
$s$, in which $a$ and $b$ are constants,
obey this smoothness condition. They are related to the limit cases of the
linear density $g(s)=a+b s$, namely, the Poisson
(zero order polynomial, $b=0$) and the
Wigner (first order polynomial with $a=0$)
distribution functions.
Let $p(s) \mbox{d} s$ denotes the probability that the next energy level 
occurs in the interval
$[e+s, e+s+\mbox{d} s]$,
\begin{equation}
p(s) \mbox{d} s = f(s) g(s) \mbox{d} s\;.
\label{defps}
\end{equation}
Then, substituting Eq.\ (\ref{fsol}) into Eq.\ (\ref{defps}), one finally
arrives at the general distribution:
\begin{equation}
p(s)=C g(s)  \exp\left(-\int_0^s g(s^\prime) \mbox{d} s^\prime\right).
\label{psol}
\end{equation}
The boundary conditions in solving the differential equation (\ref{diff})
accounts for the meaning of the NNSD
$p(s)$ and its argument as the spacing between the nearest neighbor levels
as shown in the integration limit in Eq.\ (\ref{psol}). The constant $C$
is determined by the normalization condition (\ref{normcond1})
[see Eq.\ (\ref{norm})]. We have also to care of another normalization
condition (\ref{normcond2}) for using the correct $D$ units.

\end{document}